\documentclass[twocolumn,prl,showpacs]{revtex4}
\usepackage{amsmath}
\newcommand{\be}{\begin{equation}}
\newcommand{\ee}{\end{equation}}
\begin{document}
\title{Topology and Phases in Fermionic Systems}
\author{M.~B.~Hastings}
\affiliation{Center for Nonlinear Studies and Theoretical Division,
Los Alamos National Laboratory, Los Alamos, NM, 87545}
\begin{abstract}
There can exist topological obstructions to continuously deforming
a gapped Hamiltonian for free fermions into a trivial form without
closing the gap.  These topological
obstructions are closely related to obstructions to the existence of
exponentially localized Wannier functions.
We show that by taking two copies of
a gapped, free fermionic system with complex conjugate
Hamiltonians, it is always possible to overcome
these obstructions.
This allows us to write the ground
state in matrix product form using Grassman-valued
bond variables, and show insensitivity of the ground
state density matrix to boundary conditions.  
\end{abstract}
\pacs{03.67.Mn,05.50.+q,05.30.-d}
\maketitle

The distinction between different phases of matter is one of the
most basic ideas in condensed matter and statistical
physics.  In general, following Landau, to go between two states of
different symmetry in a classical system
requires a phase transition, unless a symmetry
breaking field is turned on.
In the case of quantum systems, similar ideas about symmetry hold.
Suppose there exist two local Hamiltonians, ${\cal H}_0$ and ${\cal H}_1$,
which both
have an energy gap, $\Delta E$, between a sector of approximately
degenerate ground states and the
rest of the spectrum.  Then we ask: can one 
find a family of local Hamiltonians, ${\cal H}_s$, which depend continuously
on $s$ and interpolate between ${\cal H}_0$ and ${\cal H}_1$ such that
the gap remains open?

If so, then it is possible to go from ${\cal H}_0$ to ${\cal H}_1$ without
a quantum phase transition.  However, if so, then the technique of 
quasi-adiabatic\cite{qad}
continuation can be applied and leads, in certain cases, to a proof that
there must be a phase transition due to an obstruction.
For example
in a ferromagnetic transverse field Ising model, ${\cal H}=J\sum_{i,j \, n.n.}
S_z^i S_z^j + B \sum_i S_y^i$, one can show that
it is not possible to continue from a Hamiltonian
with $J>>B$ in the ferromagnetic phase to one with $B>>J$ in the paramagnetic
phase without a phase transition unless one violates the Ising
symmetry: one cannot have all of the ${\cal H}_s$ invariant under a global
flip of all the spins.  This is a case of a symmetry obstruction.

However, there can be even worse obstructions.  In the case of a topological
obstruction, such as in the fractional quantum Hall effect, where ${\cal H}_0$
has a multiply degenerate ground state on a torus, any path in parameter
${\cal H}_s$, if it includes only local Hamiltonians, can only produce
an exponentially small splitting in the ground state sector\cite{qad,niu} unless
the gap to the rest of the spectrum is closed.  Here, symmetry does not play
a role.  A generalization of this kind of topological obstruction
was presented in \cite{bravyi}, where it was shown that if all local operators
have exponentially small matrix elements between the ground states
for ${\cal H}_0$, then this is preserved under quasi-adiabatic continuation.

In general, we will take the ability to interpolate between two
different Hamiltonians, while preserving appropriate symmetries,
without closing the gap as a definition of what
it means to be in the same phase.  
Our main result here is to construct such a path
in parameter space for a simple but important class of system: free
fermion systems.  
These systems in some cases possess topological
obstructions to continuation back to a trivial system, where
we call a system trivial if there are no terms in the Hamiltonian
coupling different sites.  Examples of
such topological obstructions include Majorana number and Chern
number\cite{kitaev1}.  The key idea of
the construction in this paper is to evade these obstructions by
doubling the system, taking two copies which are time-reverses of
each other.  

Then, given the path in parameter space ${\cal H}_s$, for $s=0$ to $1$, between
the original Hamiltonian ${\cal H}_0$ and a trivial
Hamiltonian ${\cal H}_1$, we use quasi-adiabatic continuation to
write the ground state of the original system
in a matrix product form\cite{mps} and show
insensitivity to boundary conditions, a question naturally raised
by work in \cite{mbp} and \cite{mbp2} where similar results were
found for bosonic systems; another, less-efficient,
way to write the ground state in a matrix product form is using thermal
operators following \cite{thermal}.
Finally, in the conclusion, we discuss the relation between these
results and the existence of exponentially localized Wannier functions.

{\it Family of Hamiltonians---}
We start by some definitions of the lattice and the Hamiltonian,
then present the continuous family of Hamiltonians, and finally
give the statement and then the
proof of the first result.  Consider a lattice of $V$
sites, labeled by $i,j,...$.  
Let 
\be
{\cal H}=\sum_{i,j} \Psi_i^{\dagger}
H_{ij} \Psi_j+\sum_{ij} \Delta_{ij} \Psi^{\dagger}_i\Psi^{\dagger}_j+h.c.
\ee
be the free fermionic lattice Hamiltonian.  The Hilbert space on each
site $i$ has two possible states, and $\Psi^{\dagger}_i,\Psi_i$ are
the creation and annihilation operators.  ${\cal H}$
is a many-body operator, while $H_{ij}$ represent
the matrix elements of a Hermitian $V$-by-$V$ matrix $\hat H$
and $\Delta_{ij}$ are the matrix elements of
an antisymmetric $V$-by-$V$ matrix $\hat \Delta$.
Let ${\rm dist}(i,j)$ be a metric on the lattice.  Assume the spectrum of
${\cal H}$ has a gap $\Delta E$ between the ground state and the first
excited state.  Let $A$ be the matrix
\be
\label{mjmatrix}
A=\begin{pmatrix}
\hat H & \hat \Delta \\
\hat \Delta^{\dagger} & -\hat H^*
\end{pmatrix},
\ee
so that
\be
{\cal H}=\frac{1}{2}(\Psi,\Psi^{\dagger})^{\dagger} 
\cdot A \cdot (\Psi,\Psi^{\dagger}),
\ee
where $(\Psi,\Psi^{\dagger})$ is the $2V$ dimensional vector
$(\Psi_1,\Psi_2,...,\Psi_V,\Psi_1^{\dagger},\Psi_2^{\dagger},...,\Psi_V^{\dagger})$.
This implies that the eigenvalue of $A$ which is smallest in absolute value
is equal to the gap $\Delta E$ in absolute value.

We assume that ${\cal H}$ is local in the following sense: for some
constant $\mu>0$, and for some constant $s_1$, then
for all $i$, 
we have the bound
$2 \sum_j \exp[\mu {\rm dist}(i,j)] \sqrt{|H_{ij}|^2+|\Delta_{ij}|^2}
\leq s_1 \leq \infty$.  This implies the existence of
a Lieb-Robinson bound\cite{lr,hk,ns}: there is some velocity $v$ and
some constant $c$ which
depend only on $\mu,s_1$ such that for any operators $O_X,O_Y$ with
support on sets $X,Y$ we have
$\Vert[O_X(t),O_Y]\Vert\leq c\times |X| \Vert O_X \Vert \Vert O_Y \Vert
\exp[-\mu{\rm dist}(X,Y)] (\exp(v\mu|t|)-1)$.

We introduce a system defined on two copies of the original lattice,
with the same two-state Hilbert space on each site of each copy.
We label sites on this system by $(i,a)$, where $i$ has $V$ possible values
and $a=\uparrow,\downarrow$,
and $\Psi_{i,a}^{\dagger},\Psi_{i,a}$ are the creation and annihilation
operators.  We use the metric
${\rm dist}'((i,a),(j,b))={\rm dist}(i,j)+(1-\delta_{a,b})$.

We set ${\cal H}_0$ to be the sum of ${\cal H}$ on the first copy
and ${\cal H}^{*}$ on the second copy.
We now define the continuous family of Hamiltonians with this Hilbert space:
\be
\label{familydef}
{\cal H}_s=\sqrt{1-s^2} {\cal H}_0+
s \Delta E\sum_i \Bigl(\Psi_{i,\uparrow}^{\dagger}\Psi_{i,\downarrow}^{\dagger}+h.c.\Bigr).
\ee
The results we now show are that: {\bf (1):} for all $s$, $0\leq s\leq 1$, the
Hamiltonian ${\cal H}_s$ has a gap equal to $\Delta E$; {\bf (2):} for
all $s$, $0\leq s\leq 1$, the
Hamiltonian ${\cal H}_s$ obeys the same Lieb-Robinson
bound as ${\cal H}$, namely
$2 
(\sqrt{1-s^2} 
\sum_{j} 
\exp[\mu {\rm dist}(i,j)] \sqrt{|H_{ij}|^2+|\Delta_{ij}|^2}
+s^2 \Delta E)
\leq s_1 \leq \infty$;
{\bf (3):} the ground state of ${\cal H}_1$ is a product wavefunction
of the form
\be
|\Psi\rangle=\Bigl(\prod_{i} (\frac{1}{\sqrt{2}}
\Psi^{\dagger}_{i,\uparrow}\Psi^{\dagger}_{i,\downarrow}+\frac{1}{\sqrt{2}})\Bigr)
|0\rangle,
\ee
where $|0\rangle$ is the state where all sites are empty.
The proof of (3) is immediate.  To prove (2), note that
$\Delta E$ is bounded above by $s_1$ so that we may estimate the
operator norm of
$s\Delta E (\Psi_{i,\uparrow}^{\dagger}\Psi_{i,\downarrow}^{\dagger}+h.c.)$.

We finally consider (1).
Define the matrix $C_s$ by
\be
C_s=
\begin{pmatrix}
\sqrt{1-s^2} A &  s \Delta E \openone\\
s \Delta E \openone & -\sqrt{1-s^2} A
\end{pmatrix},
\ee
where $\openone$ is the unit $2V$-by-$2V$ matrix.

Then,
\be
{\cal H}_s=\frac{1}{2}
(\Psi_{\uparrow},\Psi_{\uparrow}^{\dagger},\Psi_{\downarrow}^{\dagger},\Psi_{\downarrow})^{\dagger}
\cdot C_s \cdot
(\Psi_{\uparrow},\Psi_{\uparrow}^{\dagger},\Psi_{\downarrow}^{\dagger},\Psi_{\downarrow}).
\ee
For each eigenvalue $\lambda$ of $A$, $C_s$ has two eigenvalues equal to
$\pm \sqrt{(1-s^2) \lambda^2+s^2 \Delta E^2}$.  Since the smallest
eigenvalue of $A$ was equal to $\Delta E$ in absolute value, the
same holds for $C_s$ and hence $(1)$ follows.

Finally, we comment on the reason for doubling the system, presenting two
examples of topological obstructions.
The first example is a one-dimensional example based
on the idea of Majorana number\cite{kitaev1}.
We consider a periodic chain of $V$ sites,
labeled $j=1...V$.  We define the operators $c_{2j-1}=\Psi_j+\Psi_j^{\dagger}$
and $c_{2j}=(\Psi_j-\Psi_j^{\dagger})/i$, thus giving Majorana
operators $c_k$ defined for $k=1...2V$.  A quadratic Hamiltonian
for a Majorana system is defined by a matrix $A$ as in Eq.~(\ref{mjmatrix}),
while the ground state $\Psi_0$ of the Hamiltonian has
correlators given by
\be
\langle \Psi_0|c_j c_k|\Psi_0 \rangle=\delta_{jk}-iB_{jk},
\ee
where the
antisymmetric matrix $B$ obeying $B^2=-1$ is given by
$B=-i{\rm sgn}(A)$.
Consider the state $\Psi_{odd}$ defined by the matrix $B_{odd}$ with
$(B_{odd})_{ij}=\delta_{i+1,j}$ if $i=1$ mod 2 and
$(B_{odd})_{ij}=\delta_{i-1,j}$ if $j=1$ mod 2.  This state is the
state in which every site is occupied:
$\langle\Psi_{odd}|\Psi_i^{\dagger}\Psi_i|\Psi_{odd}\rangle=1$.
Now consider the state $B_{even}$ defined by
$(B_{even})_{ij}=(B_{odd})_{i+1,j+1}$.  Kitaev defines a Majorana
number for any anti-symmetric matrix $B$ obeying $B^2=-1$ such
that $B_{jk}$ decays sufficiently rapidly in ${\rm dist}(j,k)$.  The Majorana
number is an integer equal to $\pm 1$ for an infinite chain,
and is close to $\pm 1$ for a finite chain depending on how the
chain length compares to the rate at which the coefficients decay.

The Majorana number has opposite signs for the states $\Psi_{odd},\Psi_{even}$.
This means that
there is no way to find a family of free fermion Hamiltonians ${\cal H}_s$ which
are local and gapped which have ground state $\Psi_{even}$ for $s=0$
and $\Psi_{odd}$ for $s=1$.  To show this, suppose that such a family
did exist.  Then, given a gapped Hamiltonian, the matrix $B_s$ has
matrix elements $(B_s)_{ij}$ decaying exponentially rapidly
in $|i-j|$ due to locality of correlation functions\cite{loc} and
hence has a well-defined Majorana number up to corrections
which tend to zero as $V$ becomes large compared to the correlation
length,
and hence the Majorana
number cannot change sign in this evolution.

However, doubling the chain evades this topological obstruction
as the Majorana number is always even for the doubled system.
The second example is in two dimension\cite{kitaev1}.  Consider
a matrix $P$ such that $P^2=P$ and such that $P^2_{jk}$ is sufficiently
rapidly decaying in ${\rm dist}(j,k)$.  Then Kitaev defines a quantity
$\nu(P)$ for such a matrix which generalizes the notion of Chern
number.  This quantity $\nu(P)$ is shown to be equal to an integer in
an infinite system, and to be close to an integer in a finite system,
the error depending on how the decay of the coefficients in ${\rm dist}(j,k)$
compares to the system size.

For any gapped local Hamiltonian of the form ${\cal H}=\Psi^{\dagger}_i
H_{ij} \Psi_j$ (so that there are no pairing terms: $\Delta=0$),
we can define a projector $P_{jk}=\langle \Psi_0| \Psi^{\dagger}_j \Psi_k |
\Psi_0 \rangle$.  Since the Hamiltonian is gapped, the correlations
in the ground state $\Psi_0$ will be short range, and hence
$P_{jk}$ will decay exponentially rapidly in ${\rm dist}(j,k)$.  Hence,
if the system size is sufficiently large compared to the correlation
length, $\nu(P)$ will be close to an integer.  A factorized wavefunction
has $\nu(P)=0$, so therefore there is no way to continue from
a gapped Hamiltonians with non-vanishing
$\nu(P)$ to a Hamiltonian with a factorized ground state without
either closing the gap or violating locality.  However, by doubling
the system, and using the complex Hamiltonian for the second of the
two copies, we evade this obstruction, since the total Chern
number of the $P$ defined by the ground state of ${\cal H}_s$
system vanishes for all $s$: at $s=0$ it cancels between the two copies.

{\it Quasi-Adiabatic Continuation and Matrix Product Ground State---}
The existence of a family of Hamiltonians (\ref{familydef}) implies that
the ground state of ${\cal H}_0$ can be represented as a matrix product state
as follows.  The results that follow are valid for arbitrary families
of local Hamiltonians ${\cal H}_s$ which have a gap for all $0\leq s\leq 1$
and which have a trivial ground state at $s=1$, so we develop them
in generality.  Let $l$ be some arbitrary length scale.
Let us write an arbitrary local
Hamiltonian ${\cal H}_s$ as $\sum_Z{\cal H}_Z(s)$,
where ${\cal H}_Z(s)$ has support only on set $Z$; in this specific
problem of free fermions, $Z$ will always be a set of one or two sites.
Using the technique of quasi-adiabatic continuation, we
can define a Hermitian operator ${\cal D}$ such that $\partial_s 
\Psi_0(s)=i{\cal D}_Z(s)+\sum_Z 
{\cal O}(|Z| \exp[-l/\xi] \sqrt{l/\xi} \Vert \partial_s H_Z(s) \Vert/\Delta E)$,
where the correlation length $\xi=2v/\Delta E+\mu$.
The operator ${\cal D}_Z(s)$ has support on the set of sites within
distance $l$ of $Z$, and has $\Vert{\cal D}_Z(s)\Vert\leq \sqrt{l/\xi}
(\Vert\partial_s H_Z(s)\Vert/\Delta E)$.  For a free fermion system,
as here, the operator ${\cal D}_Z(s)$ is a bilinear in the fermion operators.

Then, for sufficiently large $l$, we can get a good approximation
to the $\Psi_0(0)$ by 
\be
\label{aprx}
\Psi_0(0) \approx \exp[i\int_1^0 {\cal D}_{s'}{\rm d}s']\Psi_0(1),
\ee
and then we can approximate the exponential by a matrix product operator
of some given bond dimension,
so that we can approximate $\Psi_0$ as a matrix product state.
It remains only to estimate the errors involved in the particular case to
determine how large the bond dimension of the matrix product
state must be.  For (\ref{familydef}), we have $|Z|\leq 2$ and
$\Vert \partial_s H_Z(s) \Vert\leq s_1$, so for some $c_1$ which is
a numeric constant of order unity, then
for
$l=c_1\log(Vs_1/\Delta E) \xi+c_2 \xi$, Eq.~(\ref{aprx}) gives
an approximation to
$\Psi_0(0)$ with error of order $\exp[-c_2]$.
The unitary operator
$\exp[i\int_1^0 {\cal D}_{s'}{\rm d}s']$ can be regarded as
time evolution under an effective time dependent Hamiltonian ${\cal D}_{s'}$.
This effective Hamiltonian has a Lieb-Robinson group velocity which
we denote $\xi'$ which is bounded by 
a constant of order unity times $s_1 l/\Delta E$.  
Then, for the given $l$, we can approximate
$\exp[i\int_1^0 {\cal D}_{s'}{\rm d}s']\Psi_1(s)$ within error $\epsilon$
by a matrix
product operator $U_{mps}$ as follows.

This kind of approximation was discussed for one dimensional spin systems
in \cite{tobias}.
The construction of \cite{tobias} proceeds by breaking the one
dimensional system into short intervals labeled $1,2,...$,
and showing that the unitary
operator can be approximately
written as a quantum circuit: $V_{1,2}V_{2,3}...U_1 U_2....$, where
the operators $U_1,U_2,...$ act only on each interval $1,2,...$ and the
unitary operators $V_{i,i+1}$ act only on the right-half of interval $i$ and
the left half of interval $i+1$, so that $[U_i,U_j]=[V_{i,i+1},V_{j,j+1}]=0$.
This construction can be directly generalized to higher dimensions\cite{pc}.
As a next step, auxiliary bond variables are introduced to write the
operators $V_{i,i+1}$ as a sum of product of operators on interval $i$
and interval $i+1$.

In our problem, in order to decompose the operator $V_{i,i+1}$
into a sum of products of operators, we must introduce Grassman-valued
bond variables.  In order to approximation
the unitary evolution to an error $\epsilon$, we need
the number of Grassman-valued bond
variables on each bond to be logarithmically large in $V,1/\epsilon$ and
proportional to $\xi'$ (for the spin system, the bond dimension
is exponentially large in this quantity).
Thus, we can approximate $\Psi_0(0)$ by a matrix product state 
$\Psi_{mps}=
U_{mps} \Psi_0(1)$ with
Grassman-valued bond variables.
We note that the bond dimension required for this
free fermion system is much smaller than that for quasi-adiabatic
continuation of a general interacting system, where a bond dimension
polylogarithmically large in $V$ may be required, because ${\cal D}_Z$
is a fermion bilinear.

{\it Sensitivity to Boundary Conditions---}
The final question we consider is the sensitivity of such a gapped
fermion system (\ref{familydef}) to boundary conditions at $s=0$.
Consider two systems with Hamiltonians ${\cal H},{\cal H}'$ which
agree on some set of
sites $X$ in the following sense: for any operator $O$ with support
on $X$, $[O,{\cal H}]=[O,{\cal H}']$.
Now, let $Y$ be some subset of $X$ such that the set of all points $i$
with ${\rm dist}(i,Y)\leq l$ is a subset of $X$.  We then
consider the difference between the reduced density matrices, $\rho_{Y},
\rho_Y'$, of the two systems.
We will show 
that as $l$ gets large, the difference between the two density matrices
converges to zero in trace norm.  Physically, this may be viewed as
a statement about insensitivity to boundary conditions: even if two systems
differ at the ``boundary" (that is, outside $X$), far enough
away from the boundary (that is, inside $Y$ for large enough $l$) the
physical properties will be the same.

We again use quasi-adiabatic
continuation.  We take two copies of each system, and define
a continuous family of Hamiltonians ${\cal H}_s,{\cal H}'_s$ as above,
so that ${\cal H}_1={\cal H}'_1$.
Let $\Psi_0'(s),\Psi_0'(s)$ be the ground states of these Hamiltonians.
Then\cite{qad} for any operator
$O$ with support on $Y$
we can define an operator $O(s)$ such that
$O(s)$ is supported on $X$
$|\langle \Psi_0(1)|O(1)|\Psi_0(1) \rangle-
\langle \Psi_0(0)|O|\Psi_0(0) \rangle|
\leq \epsilon$ where the error $\epsilon$ is exponentially small
in $l$\cite{qad}.
Since ${\cal H}_s,{\cal H}'_s$ agree on $X$, we find that
also
$|\langle \Psi'_0(1)|O(1)|\Psi'_0(1) \rangle-
\langle \Psi'_0(0)|O|\Psi'_0(0) \rangle|
\leq \epsilon$.
However, since ${\cal H}_1={\cal H}'_1$, we have
$\langle \Psi_0(1)| O(1)|\Psi_0(1)\rangle=\langle \Psi_0'(1)|O(1)|
\Psi_0'(1)\rangle$, and so
$|\langle \Psi_0(0)|O|\Psi_0(0) \rangle
-\langle \Psi_0'(0)|O|\Psi_0'(0) \rangle|\leq 2 \epsilon$
for any $O$ with support
on $Y$.  Therefore, we can bound the trace norm distance between
$\rho_{Y,1}$ and $\rho_{Y,2}$:
${\rm tr}(|\rho_{Y,1}-\rho_{Y,2}|)\leq 2\epsilon$.
We can estimate $\epsilon$ in the specific case of a $d$-dimensional
free fermionic system to get
\be
{\rm tr}(|\rho_{Y,1}-\rho_{Y,2}|)\leq
const. \times
|Y| 
\sqrt{l/v\Delta E} s_1
\exp[-l/\xi'] (\xi')^d.
\ee

{\it Discussion and Wannier Functions---}
Can there be topological obstructions to continuing a 
{\it time-reversal symmetric} undoubled
system back to a trivial system?
One possible example is the spin Hall state\cite{spinhall}, but by
turning on an external magnetic field, breaking
the time reversal symmetry, it is possible to continuously deform the
undoubled Hamiltonian to a trivial Hamiltonian.

In order to understand possible obstructions in time reversal symmetric
systems better, we consider Wannier functions.  
We say that a system has exponentially localized Wannier functions if
one can find
a basis of states $|\phi_n\rangle$ which span the occupied states
of the free fermion system (so that the Green's function is equal
to $G=\sum_{n} |\psi_n\rangle\langle\phi_n|$) and which are exponentially
localized in real space.  Such functions have long been considered
for periodic systems, as well as more recently for disordered systems\cite{wfd}.

If a free fermion
Hamiltonian ${\cal H}_0$ can be continued to a trivial Hamiltonian
${\cal H}_1$, then
exponentially localized Wannier functions exist: the Hamiltonian
${\cal H}_1$ has Wannier functions $|\phi_n\rangle$,
where each $|\phi_n\rangle$ has support on a single site.  Then, by
quasi-adiabatically continuing from ${\cal H}_1$ to ${\cal H}_0$, we can
find exponentially localized Wannier functions for ${\cal H}_0$.
Only very recently\cite{wf} was it shown that for time reversal symmetric,
periodic systems in two and three dimensions exponentially localized
Wannier functions do exist, in that case without any doubling required.
On the other hand, the proof in \cite{wf} suggests that there exist
time-reversal symmetric,
four-dimensional, periodic systems for which exponentially
localized Wannier functions do not exist.  These systems would be
in a different topological phase from
a trivial Hamiltonian, and even by adding a magnetic field it is not
possible to continue back to the trivial Hamiltonian without closing a gap.

An interesting open question is when
such exponentially localized
Wannier functions must exist for gapped, undoubled, time-reversal
symmetric, disordered systems.  In one dimension, with open boundary
conditions, this can be shown as follows: let $X$ be the position operator.
Then, $GXG$ is a Hermitian operator and can be diagonalized; 
because $G$ is short-range due to
the gap\cite{loc}, the eigenvectors of $GXG$ are exponentially localized
in real-space 
and hence can be used as Wannier functions.  In
two dimensions, we consider the pair of operators $GXG,GYG$.  These
operators almost commute, since $\Vert [GXG,GYG] \Vert \sim O(1) <<
\Vert GXG \Vert\sim L$, where $L$ is the linear size of the system.
If these operators are are close to a pair of commuting
operators we could simultaneously diagonalize those operators and
use their eigenfunctions as Wannier functions.
The question of when almost commuting operators are
close to commuting operators is well-studied\cite{ac},
and those results can be used to show that given a family of systems of
increasing size $L$ and constant gap, then for any $\delta>0$ for sufficiently
large $L$ we can define Wannier functions localized on a scale $\delta L$.
In three dimension, we
have three almost commuting operators $GXG,GYG,GZG$,
which need not be close to commuting operators
as shown in \cite{3d}.

{\it Acknowledgments---} This work was inspired by discussions with
M. B. Plenio at the workshop on ``Lieb-Robinson Bounds and Applications" at the Erwin Schr\"{o}dinger Institute.  I thank S. Virmani for useful comments.
This work supported by U. S. DOE Contract No. DE-AC52-06NA25396.


\begin{thebibliography}{99}
\bibitem{qad} M. B. Hastings and X.-G. Wen, Phys. Rev. B {\bf 72}, 045141
(2005).

\bibitem{niu} X.-G. Wen and Q. Niu, Phys. Rev. B {\bf 41}, 9377 (1990).

\bibitem{bravyi} S. Bravyi, M. B. Hastings, and F. Verstraete, Phys.
Rev. Lett. {\bf 97}, 050401.

\bibitem{kitaev1} A. Kitaev, Ann. Phys. {\bf 321}, 2 (2006),
section c.4.2. and section c.3.1.

\bibitem{mps} M. Fannes, B. Nachtergaele, and R. F. Werner, Commun.
Math. Phys. {\bf 144}, 443 (1992).

\bibitem{mbp} M. B. Plenio and S. Virmani, Phys. Rev. Lett. {\bf 99},
120504 (2007).

\bibitem{mbp2} M. B. Plenio and S. Virmani, arXiv:0710.3299.

\bibitem{lr} E. H. Lieb and D. W. Robinson, Commun. Math. Phys. {\bf 28},
251 (1972).

\bibitem{hk} M.  B. Hastings and T. Koma,
Commun. Math. Phys. {\bf 265}, 781 (2006).

\bibitem{ns} B. Nachtergaele and R. Sims, Commun. Math. Phys.
{\bf 265}, 119 (2006).

\bibitem{loc} M. B. Hastings, Phys. Rev. Lett. {\bf 93}, 140402 (2004).

\bibitem{thermal} M. B. Hastings, Phys. Rev. B {\bf 76}, 035114 (2007).

\bibitem{tobias} T. J. Osborne, Phys. Rev. Lett., {\bf 97}, 157202 (2006).

\bibitem{pc} T. J. Osborne, private communication.

\bibitem{spinhall} C. L. Kane and E. J. Mele, Phys. Rev. Lett. {\bf 95},
226801 (2005).

\bibitem{wf} G. Panati, Ann. Henri Poincare, {\bf 8}, 995 (2007);
C. Brouder, G. Panati, M. Calandra.
C. Mourougane, and N. Marzari, Phys. Rev. Lett. {\bf 98}, 046402 (2007).

\bibitem{wfd} P. L. Silvestrelli, N. Marzari, D. Vanderbilt, and M. Parrinello,
Solid State Commun. {\bf 107}, 7 (1998).

\bibitem{ac} H. Lin, Op. Alg. and their Aplic., Fields Institute
Communications, {\bf 13}, 193 (1995); P. Friis and M. Rordam, J. Reine
Angew. Math. {\bf 479}, 121 (1996).

\bibitem{3d} K. R. Davidson, Math. Scand. {\bf 56}, 222 (1985);
D. Voiculescu, J. Op. Thy. {\bf 5}, 147 (1981).
\end{thebibliography}
\end{document}